\documentclass[11pt]{article}
\usepackage{graphicx}
\usepackage{epsfig}
\usepackage[figuresright]{rotating}
\usepackage{amsmath}
\usepackage{amssymb}
\usepackage{amsfonts}

\usepackage{cite}
\begin{document}
\begin{center}
{\LARGE{\bf Note on 
the  magnetic moments of the nucleon 
  }}\\[1ex] 
Martin Schumacher\\mschuma3@gwdg.de\\
Zweites Physikalisches Institut der Universit\"at G\"ottingen,
Friedrich-Hund-Platz 1\\ D-37077 G\"ottingen, Germany
\end{center}

\begin{abstract}
The Goldberger-Treiman relation $M=2\pi/\sqrt{3}\,f^{\rm cl}_\pi$ 
where $M$ is the
constituent quark mass in the chiral limit (cl) and $f^{\rm cl}_\pi$ the 
pion decay
constant in the chiral limit  predicts
constituent quark masses of $m_u=328.8\pm 1.1$ 
MeV and $m_d=332.3\pm 1.1$ MeV for the up 
and down quark, respectively, when $f^{\rm cl}_\pi=89.8\pm 0.3$
MeV is adopted. Treating the constituent quarks as bare
Dirac particles the following zero order values 
$\mu^{(0)}_p=2.850\pm 0.009$ and $\mu^{(0)}_n=-1.889 \pm 0.006$
are obtained for the proton  and neutron  magnetic
moments,
leading to deviations from the experimental data of 2.0\% and 1.3\%,
respectively. 
These unavoidable deviations 
are  discussed in terms of contributions to the magnetic moments
 proposed in previous work.
\end{abstract}

\section{Introduction}

The prediction of the magnetic moments of octet baryons in a constituent quark
model obeying $SU(6)$ spin-flavor symmetry
has attracted many researchers over a long period of time (see
\cite{lipkin78,lipkin81,brown80,isgur80,brown83,hong93,hong94,dannbom97,glozman99,diaz04,scadron06}
and references therein).
The overall success of these investigations has become one of the main
supports for the validity of the constituent quark model.
Furthermore, theoretical evidence has been presented that the constituent
quarks behave like bare Dirac particles \cite{weinberg90}.
 Remaining discrepancies showing up
in   previous work have been  removed in the latest of this
series of papers \cite{scadron06} where a general agreement was achieved 
between the experimental data and the predictions. One very remarkable 
result  of
this latter investigation is that the general agreement is obtained
by a proper determination of the constituent quark masses, showing that
other effects  on the magnetic moments are of minor importance.
However, by adjusting the predictions to the experimental magnetic moment
of the proton \cite{scadron06} a discrepancy between theory and experiment
of $3.3\%$ is obtained in case of the neutron.
This discrepancy shows that at a few-percent level of precision  
the proper choice of the constituent
quark mass is not sufficient for obtaining a complete agreement between theory
and experiment for both nucleons. Researches on possible additional
contributions to the octet baryon magnetic moments were carried out in
\cite{brown80,isgur80,brown83,hong93,hong94,dannbom97,glozman99,diaz04}.  
The following additional contributions were discussed:\\
(i) relativistic effects \cite{isgur80,dannbom97,diaz04},\\
(ii) configuration mixing in the ground state wave functions
\cite{isgur80},\\
(iii) loop corrections \cite{glozman99}, \\
(iv) loop and vertex corrections \cite{lvov04}, and\\
(v) pion exchange currents between constituent quarks 
\cite{brown80,brown83,hong93,hong94,dannbom97}.

The present investigation is  motivated by the fact that two other fundamental
structure constants of the nucleon, $viz.$ the electric and magnetic
polarizabilities $\alpha$ and $\beta$ have been successfully predicted on an
absolute scale by treating them as composites of the   
nucleon structure (or $s$-channel)
parts $\alpha^s$ and $\beta^s$ and the   $t$-channel parts $\alpha^t$ and
$\beta^t$,  where the $t$-channel parts could be  quantitatively  
predicted \cite{schumacher06,schumacher07a,schumacher07b,schumacher08}
on the basis of the Goldberger-Treiman relation on the quark level 
$M=2\pi/\sqrt{3}\,f^{\rm cl}_\pi$ derived by Delbourgo and Scadron 
\cite{delbourgo95} where $M$ is the mass of the constituent quark in the
chiral limit and $f^{\rm cl}_\pi$ the pion decay constant in the chiral limit.
The Goldberger-Treiman relation has been derived in a model which the authors
\cite{delbourgo95}
name the dynamically generated linear sigma model (L$\sigma$M) 
on the quark level. In our previous work  
\cite{schumacher06,schumacher07a,schumacher07b,schumacher08} 
and in the present work
no use is made of properties of the L$\sigma$M, except for the 
Goldberger-Treiman relation which has been derived from it. Our attitude
is to use the  Goldberger-Treiman relation in the form  
$M=2\pi/\sqrt{3}\,f^{\rm cl}_\pi$ independent of  the special 
method of its derivation and to find
experimental arguments which support its usefulness and validity.
The pion decay constant in the chiral limit $f^{\rm cl}_\pi$ has been derived
from the experimental pion
decay constant $f_\pi = (92.42\pm 0.26)$ MeV through a small correction
given in \cite{nagy04}. Therefore, the statement is allowed that 
the $t$-channel parts
of the electromagnetic polarizabilities are predicted on an absolute scale
 \cite{schumacher06,schumacher07a,schumacher07b,schumacher08}
with the experimentally known pion decay as the only input.
A second   available  case  is the
prediction of the two-photon width $\Gamma(\sigma\to \gamma\gamma)$ 
of the $\sigma$ meson \cite{schumacher08,scadron04}. However, in this 
latter case the
experimental value to compare with is not very precise.
As a further result  the Goldberger-Treiman relation  leads to  predictions
for the constituent quark masses on an absolute scale 
and we consider it very interesting to
investigate to what level of precision the magnetic moments of the nucleon
can be predicted on this basis.

\section{Predictions based on the
  Goldberger-Treiman relation on the quark level}

In the   dynamically generated L$\sigma$M on the quark level
which is related to the bosonized 
Nambu--Jona-Lasinio (NJL) model
the gap parameter or constituent quark mass  $M$ in the chiral limit
and the pion decay constant in the chiral limit $f^{\rm cl}_\pi$ are related to
each other through the relations \cite{schumacher06,delbourgo95}
\begin{eqnarray}
&&f^{\rm cl}_\pi=-4 i N_c\, g M \int \frac{d^4p}{(2\pi)^4}
\frac{1}{(p^2-M^2)^2},\label{constituent-1}\\
&&M=-\frac{8iN_c\,g^2}{(m^{\rm cl}_\sigma)^2}\int\frac{d^4p}{(2\pi)^4}
\frac{M}{p^2-M^2}\label{constituent-2},
\end{eqnarray}
where $N_c=3$ is the number of colors, $m^{\rm cl}_\sigma$ the mass of
the $\sigma$ meson in the chiral limit and $g$ the Yukawa coupling constant
which is related to the quantities $f^{\rm cl}_\pi$ and $M$ through the
Goldberger-Treiman relation for the chiral limit:
\begin{equation}
g f^{\rm cl}_\pi =M.
\label{constituent-3}
\end{equation}
Applying dimensional regularization in (\ref{constituent-1}) and 
(\ref{constituent-2}) and using (\ref{constituent-3})
and  $m^{\rm cl}_\sigma=2M$, we arrive at
\begin{equation}
g=g_{\pi qq}=g_{\sigma qq}=2\pi/\sqrt{N_c}=3.63.
\label{dimensional-2}
\end{equation}
The pion decay constant in the chiral limit is 
\cite{nagy04}
$f^{\rm cl}_\pi=89.8\pm 0.3$ MeV. Using this value and applying
(\ref{constituent-3}) and 
(\ref{dimensional-2}) the following value for the constituent quark 
mass in the chiral limit is obtained:
\begin{equation}
M=325.8\pm 1.1\,\, {\rm MeV}.
\label{constituentmass-1}
\end{equation}
According to the PDG \cite{PDG} the presently accepted values of the 
current quark masses are
\begin{eqnarray}
&&m^{\rm curr.}_u=3.0 \,\,{\rm MeV}, \label{currentu}\\
&&m^{\rm curr.}_d=6.5 \,\,{\rm MeV}. \label{currentd}
\end{eqnarray}
This leads to the predicted constituent quark masses
\begin{eqnarray}
&&m_u=M+m^{\rm curr.}_u=328.8\pm 1.1\,\,{\rm MeV}, \label{conmassuu}\\
&&m_d=M+m^{\rm curr.}_d=332.3\pm 1.1\,\,{\rm MeV}. \label{conmassdd}
\end{eqnarray}

The argument leading to (\ref{conmassuu}) and (\ref{conmassdd})
may be found in Eq. (4.19) of \cite{klevansky92} where a gap equation is
formulated for the constituent quark mass $m^*$ including 
the effects of explicit symmetry breaking. This gap equation Eq. (4.19)
shows that (\ref{conmassuu}) and (\ref{conmassdd}) are  valid except for
a very small and, therefore, negligible correction.

In principle the quantities $g$ and $M$ in Eqs. (\ref{dimensional-2}) and
(\ref{constituentmass-1}) may depend on the regularization scheme. 
However, as has been shown already in a previous work \cite{schumacher06}  
this dependence on the regularization scheme apparently is marginal. 
The argument 
was as follows. When we calculate the mass of the $\sigma$ meson according to
\begin{equation}
m_\sigma=(4M^2+m_\pi^2)^{1/2}
\label{sigmamass-1}
\end{equation}
(see e.g. \cite{klevansky92}) we arrive at
\begin{equation}
m_\sigma=666.0\,\,{\rm MeV}.
\label{sigmamass-2}
\end{equation}
An independent  calculation \cite{hatsuda94} (see also the discussion
in \cite{schumacher06}) in terms of the four-fermion
version of the NJL model with regularization through a cut-off parameters
$\Lambda$  has led to 
\begin{equation}
m_\sigma\simeq668\,\,{\rm MeV}.
\label{sigmamass-3}
\end{equation}
The good agreement of the numbers in  (\ref{sigmamass-2})
and (\ref{sigmamass-3}) gives us confidence that the 
dependence of the quantity $M$ and
consequently also of the constituent quark masses $m_u$ and $m_d$ given in 
(\ref{conmassuu}) and (\ref{conmassdd}) on the regularization scheme is
very small.

The spin-dependent part of the nucleon wave function may be given in the form
\begin{eqnarray}
|p\rangle=\sqrt{2/3}\,\chi(1,1)\,\phi(1/2,-1/2)-
\sqrt{1/3}\,\chi(1,0)\,\phi(1/2,1/2),
\label{magneticmoment-1}\\
|n\rangle=\sqrt{2/3}\,\phi(1,1)\chi(1/2,-1/2)-
\sqrt{1/3}\,\phi(1,0)\,\chi(1/2,1/2).\label{magneticmoment-2}
\end{eqnarray}
where $\chi(J,M)$ represents the up quarks and $\phi(J,M)$ the down
quarks. This leads to the magnetic moments
\begin{eqnarray}
\mu_p=\frac23(2\mu_u-\mu_d)+\frac13 \mu_d=\frac43\mu_u-\frac13 \mu_d, 
\label{magneticmoment-3}\\
\mu_n=\frac23(2\mu_d-\mu_u)+\frac13 \mu_u=\frac43\mu_d-\frac13 \mu_u,
\label{magneticmoment-4}
\end{eqnarray}
in units of the nuclear magneton  $\mu_N=e\hbar/2m_p$ \cite{PDG}. 
Constituent quark masses enter  through the relations
\begin{equation}
\mu_u=\frac23\, \frac{m_p}{m_u},\quad 
\mu_d=-\frac13\, \frac{m_p}{m_d}.
\label{intro-2a}
\end{equation}

Using the constituent quark masses given in (\ref{conmassuu}) and
(\ref{conmassdd}) the  zero-order
values of the magnetic moments of the nucleon
can be calculated via 
\begin{equation}
\mu^{\rm (0)}_p=\frac13 \left(4\mu^{\rm (0)}_u-\mu^{\rm (0)}_d \right),
\quad \mu^{\rm (0)}_n=\frac13 \left(4\mu^{\rm (0)}_d-\mu^{\rm (0)}_u 
\right),\label{quarkano-1}
\end{equation}
with
\begin{equation}
\mu^{\rm (0)}_u=\frac23 \frac{m_p}{m_u}=1.902,\quad  
\mu^{\rm (0)}_d=-\frac13 \frac{m_p}{m_d}=-0.941.
\label{quarkano-2}
\end{equation}
This leads to the results given in Table \ref{tableresults-1}. In addition
to the data for the proton and the neutron also the isoscalar 
$\mu_S=\frac12 (\mu_p+\mu_n)$ and isovector magnetic moments
$\mu_V=\frac12 (\mu_p-\mu_n)$
are given.
\begin{table}[h]
\caption{Predicted  magnetic moments of the nucleon in zero-order
approximation $\mu^{(0)}$ compared with 
experimental data. The quantities $\mu^{(0)}$ have been calculated from
Eqs. (\ref{quarkano-1}) and (\ref{quarkano-2}). 
The corrections $\mu^{\rm corr.(0)}$ 
are  the differences between the experimental
values $\mu^{\rm exp.}$ and the zero-order  values  $\mu^{(0)}$.} 
\begin{center}
\begin{tabular}{lllll}
& proton & neutron & isoscalar & isovector\\
\hline
$\mu^{\rm (0)}$  & $+2.850\pm 0.009$ & $-1.889\pm 
0.006$ & $ +0.480$& $+2.370$\\
$\mu^{\rm exp.} $   & $+2.793 $ & $ -1.913$ & $+0.440$ & $+2.353$ \\
\hline
$\mu^{\rm corr.(0)}$& $-0.057\pm 0.009$ & $ -0.024\pm 0.006$ 
& $-0.040$& $-0.017$ \\
\hline
\end{tabular}
\end{center}
\label{tableresults-1}
\end{table}
It is
interesting to note that the difference between the zero-order values 
$\mu^{(0)}$ and experimental values   $\mu^{\rm exp.}$
amounts to only $ 2.0\%$ for the proton and  $ 1.3\%$ for the neutron.
This difference is mainly isoscalar.

It may be of interest to compare the predictions of the Goldberger-Treiman
relation with an  approach 
where the constituent quark masses are adjusted
to the magnetic moments of the nucleon using 
(\ref{magneticmoment-3}), (\ref{magneticmoment-4}) and (\ref{intro-2a})
as has been
done in previous work \cite{scadron06,scadron07}.
 Then  the
quantity $M$ entering into (\ref{conmassuu}) and  (\ref{conmassdd}) is an
adjustable parameter which may be denoted by $M(p,n)$. 
In this case the constituent-quark masses are
\begin{equation}
m^{(p)}_u=335.6\,\,\, {\rm MeV},\quad m^{(p)}_d=339.1\,\,\,{\rm MeV} ,
\label{adjust-1}
\end{equation}
when adjusted to the magnetic moment of the proton and 
\begin{equation}
m^{(n)}_u=324.6\,\,\, {\rm MeV},\quad m^{(n)}_d=328.1\,\,\, {\rm MeV},
\label{adjust-2}
\end{equation}
when adjusted to the magnetic moment of the neutron.
An interesting feature of this prediction of the constituent quark
masses is that the arithmetic averages 
\begin{equation}
\frac12(m^{(p)}_u+m^{(n)}_u)= 330.1\,\,\,{\rm MeV}\quad{\rm and}\quad
 \frac12(m^{(p)}_d+m^{(n)}_d)= 333.6\,\,\,{\rm MeV}
\label{adjustedmass}
\end{equation}
both are larger than the corresponding predictions of the Goldberger-Treiman 
relation in (\ref{conmassuu}) and (\ref{conmassdd}) by $+0.4\%$. Therefore,
this
difference between the results in   
(\ref{conmassuu}) and (\ref{conmassdd}) and in (\ref{adjustedmass})
may be interpreted in terms of the  uncertainty of the pion decay constant
$f^{\rm cl}_\pi$. First we notice that by inserting
$M(p,n)=327.1$ MeV instead of $M=325.8$ MeV into 
(\ref{conmassuu}) and (\ref{conmassdd}) we exactly arrive at the numbers given
(\ref{adjustedmass}). This means that the  
use of  $M(p,n)=327.1$ MeV instead of $M=325.8$ MeV
may be understood in terms of a shift
\begin{equation}
f^{\rm cl}_\pi=89.8\,\,\, {\rm MeV} 
\Longrightarrow f^{\rm cl}_\pi=90.1\,\,\, {\rm MeV}
\label{shift}
\end{equation}
of the pion decay constant.
This shift by $\sim 0.3\%$ is within the error of the quantity 
$f^{\rm cl}_\pi$.

\section{Discussion}

The surprising feature of the numbers in Table  \ref{tableresults-1} 
is that 
the correction terms $\mu^{\rm corr.(0)}$    are so small. This is quite 
satisfactory because it gives us a further good example 
that predictions  obtained on the basis
of the Goldberger-Treiman relation on the quark level are valid to a 
high level of precision. 
Nevertheless it appears justified to ask for
reasons that these correction terms exist. For this purpose we discuss
one of the previous proposals which at  first sight 
appears to us especially relevant and which remains within the present ansatz
of a bare Dirac particle. This is the configuration mixing.

In the $SU(6)$ harmonic oscillator basis the ground state of the nucleon
may be given in the form
\begin{equation}
|P_{11}(939)\rangle=a_S|N\, ^2S_{1/2}\rangle_S + a'_S|N\, ^2S'_{1/2}\rangle_S
+a_M|N\,^2S_{1/2}\rangle_M+a_D|N\,^4D_{1/2}\rangle_M,
\label{wavefunction}
\end{equation}
where the coefficients have been determined  \cite{giannini90} to be
\begin{equation}
a_S=0.931,\quad a'_S=-0.274,\quad a_M=-0.233,\quad a_D=-0.067.
\label{coefficients}
\end{equation}
The first two terms on the r.h.s. of (\ref{wavefunction}) differ by the
oscillator quantum number $N$, being $N=0$ and $N=2$ respectively,
but have the same $SU(6)$ structure otherwise. The $D$ wave admixture
represented by the last term enters with a coefficient of $P_D=a^2_d=0.0045$
into the expression for the magnetic moment and therefore may be disregarded.
This justifies to treat the nucleon ground state as a linear combination
of only $^2S_S$ and $^2S_M$ components, so that the magnetic moments can be
expressed as follows \cite{isgur80}:
\begin{eqnarray}
&&\mu^{\rm conf.}_p=\frac13(4\mu_u - \mu_d)\, \cos^2\phi^N_8 
+\frac13(2\mu_u+\mu_d)\, \sin^2\phi^N_8,\label{isg-1}             \\
&&\mu^{\rm conf.}_n=\frac13(4\mu_d - \mu_u)\, \cos^2\phi^N_8 
+\frac13(2\mu_d+\mu_u)\, \sin^2\phi^N_8.\label{isg-2}
\label{confmix}
\end{eqnarray}
The component $^2S_S$ corresponds to the quark structure given in 
(\ref{magneticmoment-1}) and (\ref{magneticmoment-2})
or to $[56,0^+]$ states in $SU(6)$ notation  whereas the impurity
$^2S_M$ corresponds to $[70,0^+]$ states in $SU(6)$ notation. These impurities
have been  introduced as a consequence of color hyperfine interactions.
In  \cite{isgur80} the mixing amplitude is given as $\sin \phi^N_8= -0.27$
in close agreement with $a_M=-0.233$.  Using $\sin \phi^N_8= -0.27$
and the zero-order predictions for the quark magnetic moments given
in  (\ref{quarkano-2}) we arrive at corrections due to configuration mixing as
given in Table 2.
\begin{table}[h]
\caption{Predicted magnetic moments of the nucleon including 
configuration mixing  $\mu^{\rm conf.}$ calculated from
Eqs. (\ref{quarkano-2}), (\ref{isg-1}) and (\ref{isg-2}). 
The quantity $\mu^{\rm corr.(conf.)}$ is the difference between 
$\mu^{\rm exp.} $ and $\mu^{\rm conf.}$.
}
\begin{center}
\begin{tabular}{lllll}
& proton & neutron & isoscalar & isovector\\
\hline
$\mu^{\rm conf.}$  & $+2.711$ & $-1.750$ & $ +0.480$& $+2.230$\\
$\mu^{\rm exp.} $   & $+2.793 $ & $ -1.913$ & $+0.440$ & $+2.353$ \\
\hline
$\mu^{\rm corr.(conf.)}$& $+0.082$ & $ -0.163$ & $-0.040$& $+0.123$ \\
\hline
\end{tabular}
\end{center}
\label{tableresults-2}
\end{table}
In Table \ref{tableresults-2} the discrepancy between experiment and prediction
is 3\% for the proton and 9.3\% for the neutron. Apparently, 
 the discrepancies are much larger when the configuration mixing is included 
than in  case of the zero-order
predictions  $\mu^{(0)}$. This means that when introducing corrections due to 
configuration mixing  it would be necessary to simultaneously find an
other sizable effect which compensates for the configuration mixing
effects. Without going into details here the same conclusions can be
drawn for the other corrections proposed in the literature.

\section{Summary and conclusions}

It has been shown that the  magnetic moments of the nucleon can
be calculated with a high level of precision of 1--2\% on an absolute scale
using the constituent quark masses predicted on the basis of 
Goldberger-Treiman relation $M=2\pi/\sqrt{3}\,f^{\rm cl}_\pi$ 
with  $f^{\rm cl}_\pi= 89.8\pm 0.3$ MeV derived from the experimentally 
known pion decay constant $f_\pi=
(92.42\pm 0.26)$ MeV as the only input. 
The importance of the present finding is that in addition to the $t$-channel
parts of the electromagnetic polarizabilities and 
the two-photon width $\Gamma(\sigma\to \gamma\gamma)$ 
of the $\sigma$ meson \cite{schumacher08,scadron04}
we now have a further example 
where this relation is successful in predicting the correct results.
This leads us to the conclusion that the 
Goldberger-Treiman relation $M=2\pi/\sqrt{3}\,f^{\rm cl}_\pi$ predicts the
mass $M$ of the constituent quark in the chiral limit 
with a high level  of precision. Furthermore, since this prediction is
successful in connection with three different experimentally known
observables it may be concluded that this success cannot be fortuitous
but may be considered as a proof for the general validity 
of the Goldberger-Treiman relation.

 Another important results
is that none of the available predictions of possible deviations from
the  constituent quark approach in zero-order    approximation 
with the constituent quarks treated as bare Dirac particles
leads to an explanation of the
corrections terms $\mu^{{\rm corr.} (0)}$ given in Table \ref{tableresults-1}. 
This has been explicitly shown for configuration mixing  
but is also true for the other cases listed in the introduction.
Therefore, the  explanation of this
residual discrepancy remains a problem for future work.

\section*{Acknowledgment}

This note is an extended version of a short communication presented by the
present author in a discussion on the conference Scadron 70  Workshop on
``Scalar Mesons and Related Topics'', February 11--16 (2008) in Lisbon,
Portugal.

\end{document}